# Nonlinear Chiroptical Effect in Refractory Plasmonic Molybdenum Metasurface: Chiral Saturated Absorption and Chiral Reverse Saturable Absorption

Peng Yu[1,2], Tianji Liu[3,4], Yuxuan Zhu[1,2], Feng Lin[5], Shuai Yue[6], Junichi Takahara[7,8], Tao Ding[9], Hongxing Xu[10], Alexander Govorov[11],* Zhiming Wang[1,2]*

[1]Institute of Fundamental and Frontier Sciences, University of Electronic Science and Technology of China, Chengdu 610054, China

[2]Shimmer Center, Tianfu Jiangxi Laboratory, Chengdu 641419, P. R. China

[3]GPL Photonics Laboratory, State Key Laboratory of Luminescence Science and Technology, Changchun Institute of Optics, Fine Mechanics and Physics, Chinese Academy of Sciences, Changchun 130033, China

[4]University of Chinese Academy of Sciences, Beijing 100039, China

[5]National Center for International Research on Photoelectric and Energy Materials, School of Materials and Energy, Yunnan University, Kunming, Yunnan 650091, PR China

[6]CAS Key Laboratory of Standardization and Measurement for Nanotechnology, National Center for Nanoscience and Technology, Beijing 100190, P.R. China

[7]Graduate School of Engineering, Osaka University, 2-1 Yamadaoka, Suita, Osaka 565-0871, Japan

[8]Photonics Center, Graduate School of Engineering, Osaka University, 2-1 Yamadaoka, Suita, Osaka 565-0871, Japan

[9]Key Laboratory of Artificial Micro/Nano Structure of Ministry of Education, School of Physics and Technology, Wuhan University, Wuhan, 430072, China

[10]School of Physics and Technology, Center for Nanoscience and Nanotechnology, Wuhan University, Wuhan 430072, China

[11]Department of Physics and Astronomy and Nanoscale and Quantum Phenomena Institute, Ohio University, Athens, Ohio 45701, USA

peng.yu@uestc.edu.cn

govorov@ohio.edu

zhmwang@uestc.edu.cn

## Abstract

Nonlinear chiral optical responses offer significantly enhanced selectivity and sensitivity to chiral signals compared to their linear counterparts, making them highly promising for molecular detection and information processing applications. However, the high-intensity laser required for nonlinear optical processes induces substantial thermal effects, compromising the stability and sustainability of nonlinear chiroptical responses. Here, we present a refractory chiral Mo metasurface that demonstrates significant chiroptical effect even under extreme conditions, withstanding temperatures up to 1100 °C and laser intensities above 12 GW/cm². The absorptive chiral metasurface shows evidence of giant nonlinear chiroptical effects, including chiral third harmonic generation, chiral saturated absorption, and chiral reverse saturable absorption. We finally demonstrate proof of concept circular polarized light limiter based on the refractory chiral Mo metasurface and paves the way for nonlinear chiral sensing, nonlinear hot electron generation, and XOR gate of the all-optical full-adder, showcasing the potential of refractory chiral Mo metasurfaces in overcoming the limitations of traditional chiral plasmonic materials.

## Introduction

Chirality exists in objects that cannot be superimposed with their mirror images, and they exhibit a different response to circularly polarized light (CPL)—left−circularly polarized (LCP) and right−circularly polarized (RCP) light.[1-4] From the microscopic level, such as DNA, essential amino acids, and proteins, to the macroscopic level, such as germination, mantis shrimps, and Scarabaeidae, responses to CPL are different.[5,6] Differential absorption in the chiral medium can be probed by circular dichroism (CD, CD = $A_{LCP}-A_{RCP}$, $A$ is absorption) spectrometer due to the difference of the imaginary part of the refractive index under CPL light. The chiral efficiency is indicated by the g-factor:

$$g_{CD,optical} = 2 \cdot \frac{A_{LCP} - A_{RCP}}{A_{LCP} + A_{RCP}}$$

The g-factor for natural substances is typically in the range of about $10^{-5}$ to $10^{-7}$, mainly due to the relatively minor spatial spread of molecular wavefunctions compared to the wavelength of incoming light. The artificially designed plasmonic metamaterials and nanocrystals have empowered giant optical activities due to the plasmon-enhanced light−matter interactions. The g-factor of chiral plasmonic gold nanocrystals is in the range of 0.01–0.2 while the planar chiral metasurfaces can demonstrate g-factors above 1,[7-9] and thus demonstrate promise for enantioselective sensing,[10] image classification,[11] CPL photodetection,[12,13] fluid convection,[7,14] photochemistry and chiral growth.[15-18] However, it is difficult to achieve a large g-factor (limit to 2) in the linear optic region experimentally due to structural non-idealities, material losses, environmental interference, and limitations in fabrication and measurement technologies.

Plasmonics provide enhanced nonlinear effects with ultrafast response time and field enhancement and afford nonlinear optical components with reduced size.[19] Nonlinear chiroptical effects have demonstrated enhanced chiral sensing because of its sensitivity to the symmetry of physical interfaces and analyte molecules.[20,21] Under optical excitation $E$, the material polarization $P$ is described by:

$$P = \varepsilon_0[\chi^{(1)}E + \chi^{(2)}E^2 + \chi^{(3)}E^3 + \cdots]$$

where the $\varepsilon_0$ is the vacuum permittivity and $\chi^{(n)}$ is the *n*th-order susceptibility. The most important nonlinear applications occur in the second and third orders. Interactions between CPL and chiral nanostructure in the nonlinear domain exhibit heightened sensitivity to slight asymmetry, leading to a nonlinear optical CD that is tenfold greater compared to the signal derived at the fundamental frequency.[22-24] The nonlinear optical CD in plasmonic chiral metamaterials and metasurfaces, in particular, for second harmonic generation (SHG) and third harmonic generation (THG), has been used for chiral spectroscopy due to its sensitivity to the symmetry of molecules and nanostructures.[22,25-27] Figs. 1 outlines the development of chiral metasurfaces, including plasmonic and dielectric metastructures. Dielectric chiral metasurfaces can withstand higher laser intensities to some extent. In contrast, conventional plasmonic structures made of materials like gold and silver cannot endure high-intensity lasers, making it challenging to operate in a nonlinear regime. Strong fields must be used to excite nonlinear CD effects; however, intense laser exposure can heat chiral plasmonic structures, leading to permanent thermal deformation. Due to the size effect on melting point, conventional plasmonic nanostructures, such as gold and silver, exhibit significantly lower melting points than their bulk counterparts, leading to a low damage threshold under strong laser excitation and further limiting the conversion efficiency.[28,29] For instance, thermal induced by high-intensity laser gives rise to resonant wavelength shift and even impaired absorption spectrum.[28,30] Therefore, plasmonic metasurfaces capable of operating in the nonlinear regime are crucial for chiroptics and their applications.

In this work, we first report the experimental realization of chiral metasurface using molybdenum (Mo) with a bulk melting point of ~2623°C. Also, it exhibits a plasmonic resonance in the visible-NIR range, enabling absorption difference between LCP and RCP light illuminance if the metasurface is chiral. Our refractory chiral metasurface (RCM) is a periodic array of chiral nanostructure, with a unit cell consisting of a 'Z'-shaped Mo antenna. We use a SiO$_2$ spacer and an optically thick Mo backplane to enhance its absorption, as shown in Fig. 1a. We observe selective absorption under CPL light, indicating a large CD across the wavelength range from 1000 nm to 1600 nm. Further, we investigated the thermal stability of this chiral metasurface, which demonstrated significant differential absorption even at temperatures as high as 1100 °C and can withstand 12 GW/cm$^2$ laser illumination. Thus, it can be expanded to allow a nonlinear plasmonic effect under a high-intensity CPL pump, including nonlinear THG-CD; we first observe chiral saturable and reverse saturable absorption. The THG-CD and nonlinear g-factor of the Mo RCM approach the theoretical limit of the chiroptical effect. We further demonstrate a circular polarized light limiter based on the nonlinear effect of the RCM. This device provides an efficient platform for optoelectronic and photonic applications such as nonlinear hot electron generation, thermophotovoltaics, nonlinear chiral sensors, and plasmonic modulators.

**Structure design and electromagnetic modeling of the RCM**

The schematic of the right-handed (RH) chiral metasurface is shown in Fig. 1a, and the unit cells consist of two connected Mo rectangles with overlapped space *s*. They have a constant height of 55 nm, width *w* 100 nm, and varying length *l* on top of a 140 nm SiO$_2$ substrate, backed by a 200 nm Mo film, as shown in Fig. 1b and 1c. The rectangular unit cell has periods of P1 and P2. By merging two rectangular resonators and altering their positions, the upper gold layer is endowed with simple dual-rectangle configurations that exhibit disrupted mirror symmetries, thereby granting the engineered metasurface absorber chiral properties.

We used the finite element method COMSOL multiphysics software to simulate the chiroptical response of the Mo chiral metasurface (see methods section for details). Illustrations of electromagnetic field patterns at resonant frequencies reveal the fundamental mechanisms at play. The contrast in simulated electric field patterns and surface charge distributions between LCP and RCP light is evident in Fig. 1d. The pronounced asymmetry observed at 1260 nm stems from the destructive LCP and constructive RCP interference of the incident light. One can understand the phenomena by decomposing the CPL into two orthogonal linearly polarized light with electric field vectors that exhibit a phase difference of 90 degrees. (Supplementary Figs. 2) Enhancement of the magnetic field is depicted in Supplementary Figs. 2a. Therefore, the absorption loss is concentrated in the upper resonator, depicted in Supplementary Figs. 2b, stemming from the enhanced fields around the resonators. Furthermore, Fig. 1d displays the surface charge distributions on the chiral metasurface at 1260 nm under LCP and RCP light, respectively. The profiles of surface charge distribution vary under CPL, and the electric field profile is depicted in Fig. 1d. Given the nanostructure's intricate design, the optical excitation character within the system exhibits a multipolar resonance. The discussion applies to structures with $l$ = 470, 500 nm and to their LH enantiomers and vice versa, and their corresponding electric/magnet field and surface charge distribution are shown in Supplementary Figs. 4-7.

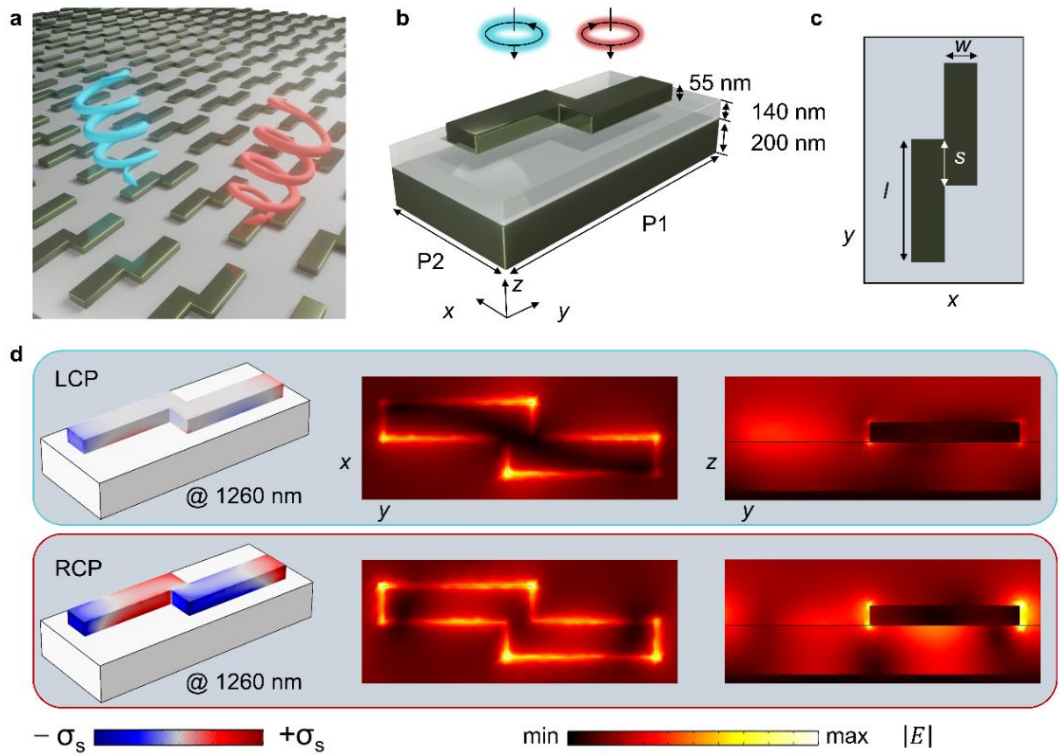

Fig. 1 **Structure design and electromagnetic simulation**. **a**, The device schematic shows the refractory Mo metasurface with chiral nanoantennas atop. **b**, **c**, The geometry of a unit cell in the chiral metasurface. The thicknesses of the top Mo chiral structure and bottom Mo layer are 55 and 200 nm, respectively; the thickness of the $SiO_2$ spacer is 140 nm; the overlapped space $s$ is set to be 70 nm; $w$ is set to be 100 nm; the unit cell has periods of P1 and P2, corresponding to different $l$. **d**, Surface charge distribution and near field properties (at 1260 nm) of the RH ($l$=420 nm, P1=890 nm, P2=350 nm) enantiomer under LCP light illumination; surface charge distribution and near field properties (at 1260nm) of the RH enantiomer ($l$=420 nm, P1=890 nm, P2=350 nm) under RCP light illumination; the surrounding medium is air; the unit of the surface charge and electric field are $C/m^2$ and V/m, respectively. The $E_z$ field is monitored along the z-axis (z=360 nm), and the $E_x$ field is monitored along the x-axis (x=150 nm)

**Fabrication of the RCM and chiroptical properties**

Experimentally, a 200 nm Mo film and 140 nm $SiO_2$ were subsequently deposited on a silicon substrate by magnetron sputtering and electron beam evaporation, respectively. Afterward, we adopt a highly standard lift-off approach to fabricate the RCM (see methods section for details). The refractive index of Mo permits plasmonic resonance in the visible to NIR range, as shown in Supplementary Figs 8. Top-down scanning electron microscope (SEM) images are shown in Fig. 2a and Supplementary Figs 9. The calculated optical absorption of RH metasurfaces with three different $l$ is shown in Fig. 2b. The absorption difference under LCP and RCP light can be observed. For the RH metasurface with $l$=420 nm, there is a resonance leading to near unity absorption of RCP light at 1260 nm while LCP light is largely reflected. As the $l$ increases, the samples show a continuous increase of the resonant wavelength from 1.26μm to 1.45μm. The significant difference in optical absorption for RCP and LCP light leads to a large CD, as shown in Fig. 2c. The calculated peak absorption achieved approaches 0.96, and the absorptive circular dichroism peaks ~0.5. As a result, the chiral efficiency g-factor is enhanced because it is dictated by the large CD, as shown in Fig. 2d. These characters indicate that the chiral metasurface can be used for chiral optics.

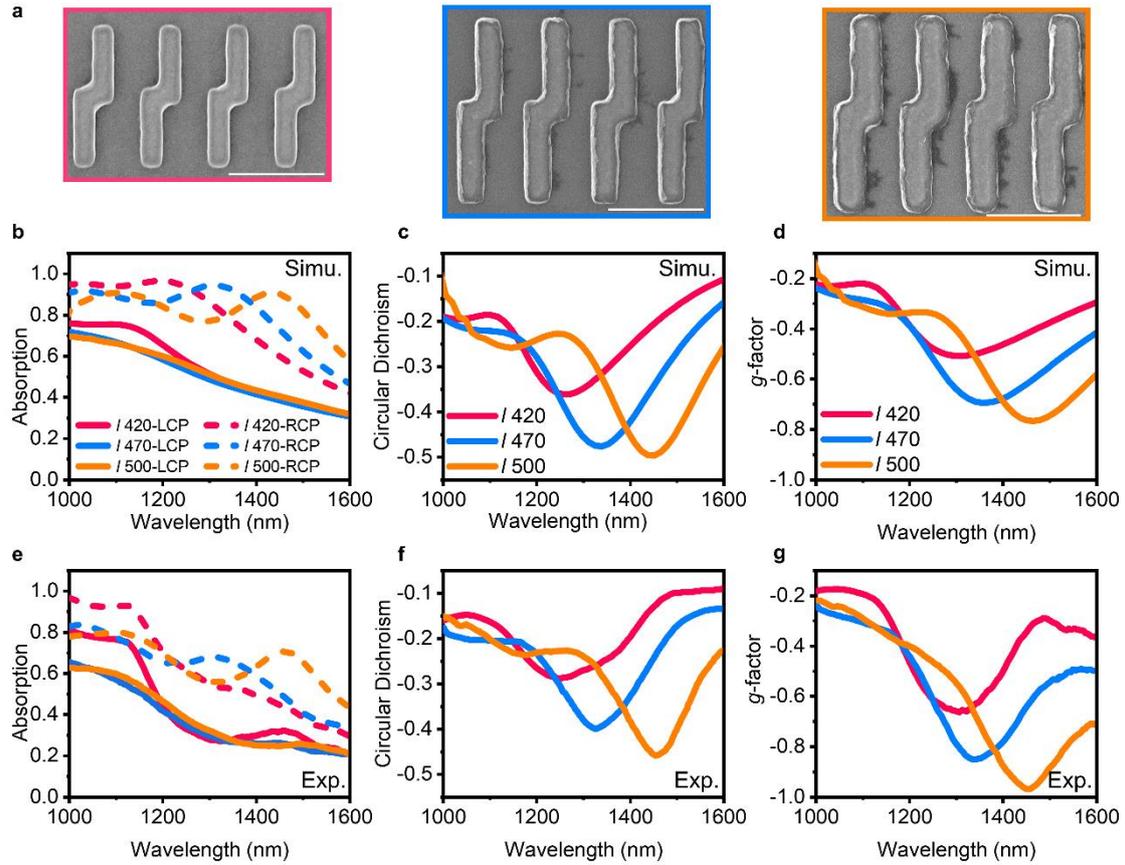

**Fig. 2 RCM and their chiroptical properties**. **a**, SEM image of three chiral metasurfaces with different *l*. Scale bar, 500 nm. **b-d**, The calculated chiroptical properties of three chiral metasurfaces with different *l*: absorption under RCP and LCP (**b**), absorptive CD (**c**), and chiral efficiency g-factor (**d**). **e-g**, The measured chiroptical properties of three chiral metasurfaces with different *l*: absorption under RCP and LCP (**e**), absorptive CD (**f**), and chiral efficiency g-factor (**g**). The metasurfaces measured have overall areas of 300 × 300 μm$^2$.

The g-factor for natural ingredients typically ranges from $10^{-7}$ to $10^{-5}$, attributed to the molecular wave functions' minimal spatial spread relative to the light's wavelength.[15] Conversely, artificial plasmonic nanostructures exhibit significantly elevated g-factors due to plasmon-enhanced light-matter interactions.[7,14,15] There is some deviation between theory and experiment, but considering factors such as the defects and roughness arising from the fabrication, the optical absorption spectra simulation results of the experimental measurement are in very good agreement (Fig. 2e-f). The chiral resonant wavelength can be tailored by altering length *l*, as demonstrated by Fig. 2b-g.

**Thermal stability properties**

For absorptive metasurfaces, the plasmonic effect enhances their light absorption with an enhanced electric field, ultimately dissipated into heat.[31,32] Herein, we investigated the thermal stability of the RCM by annealing it under elevated temperature for 4 hours. The RCM ($l$=420 nm) is heated at an increasing temperature from room temperature (RT) up to 1200 °C. After each step, the absorption spectra and SEM images are recorded. As Fig. 3a shows, the RCM has no obvious deformation until 1100 °C; at 1200 degrees °C, the heat destroys the structure. The corresponding CPL light absorption of the RCM is shown in Fig. 3b and c. For the Mo RCM, the LCP and RCP light absorption still survive up to 1100 °C due to unchanged structure shape, allowing unaltered CD and g-factor (Fig. 3d and e). Its CD center wavelength and maximum CD from RT to 1100 °C have no obvious shift, indicating it is an ideal candidate for nonlinear chiral plasmonics. For comparison, we also fabricate a chiral metasurface using conventional plasmonic gold, as shown in Supplementary Figs. 10-12. The gold chiral metasurface can withstand 400 °C, but the structure is damaged at 600 °C for only 30 minutes of heat treatment; the gold chiral nanostructures melt into large irregular structures. We also observe pitting of the gold layer result from heat at 400 °C (Supplementary Figs. 12). Compared with gold, the high thermal stability of Mo originates from its intrinsic properties, such as higher melting point, stronger bonding forces between atoms, and lower thermal expansion coefficient. Also, Mo has a body-centered cubic (BBC) structure, while gold has a face-centered cubic structure. The BCC structure generally provides higher thermal stability at high temperatures.

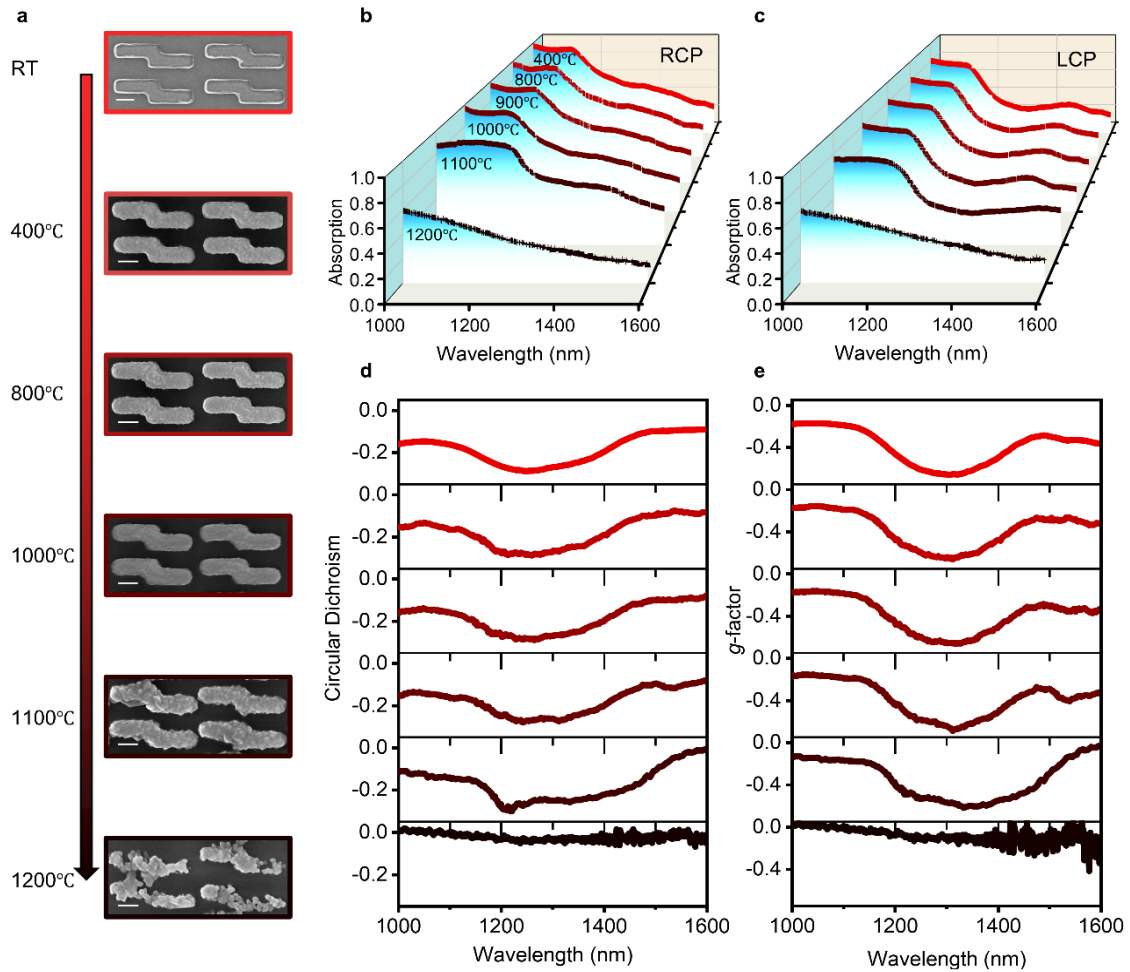

**Fig. 3 Thermal stability of the RCM. a**, SEM images of the RCM with *l*=420 nm after annealing in a vacuum at different temperatures for 4 hours. **b-e**, The measured chiroptical properties of the RCM after 4 hours of annealing at different temperatures: (**b**) RCP light absorption; (**c**) LCP light absorption; (**d**) CD spectra; (**e**) g-factor.

The thermal durability of the RCM, which is crucial for maintaining its performance over the long term at high temperatures, is of paramount importance alongside its thermal stability. Prolonged heating tests are performed at 1000 °C under vacuum conditions to assess the long-term stability of the RCM, as shown in (Supplementary Figs. 13a). Its corresponding absorption under CPL illumination, CD, and g-factor are shown in Supplementary Figs. 13b-e. The results unequivocally demonstrate that both the morphology and chiroptical properties of the RCM are well preserved even after prolonged heating, further substantiating its exceptional thermal stability. An intense laser must be used to excite the nonlinear effect in plasmonic structures. Due to the optical absorption of plasmonic structures, intense lasers can cause significant photothermal effects in metallic plasmonic structures, including localized temperature increases, leading to thermal deformation of the structure and, in extreme cases, structural damage, such as through thermal expansion or melting. The high thermal stability suggests our RCM can be used for nonlinear chiroptics due to an increased damage threshold for laser illumination. To stress our structure's excellent optical and thermal stability, we further compare our RCM with other state-of-the-art plasmonic structures for high-temperature applications, as shown in Table S1.

**Nonlinear Chiroptic Properties: THG-CD, nonlinear g-factor, and chiral saturated absorption and reverse saturable absorption**

To date, our discussion on the linear light absorption and thermal stability of the RCM has concentrated on chiroptical properties. We will now explore its nonlinear chiroptical properties under high laser power, which the THG-CD quantifies. We chose THG-CD due to its universality in all materials[19]; also, the THG signal depends on the structure shape of nanostructures crucially.[28] For metallic nanostructures, the threshold to excite nonlinear effects is typically above~1GW/cm$^2$.[33] The RCM here is exposed to 12 GW/cm$^2$ of laser power. The normalized THG-CD intensity at different times is shown in Fig. 4a. Under the illumination of RCP and LCP fundamental light, the RCM structure exhibits different THG signal intensities that remain stable over time. In contrast, the THG signal intensity of chiral gold structures quickly diminishes under the same field intensity. This demonstrates that RCM is an excellent nonlinear chiral optical material. Fig. 4b shows a microscopy image of the THG-CD signal. As expected, the RCM emits a strong THG signal under RCP light, while under LCP light, it emits a poor THG signal. Fig. 4c and d show simulated and measured THG signals. Under the pumping of RCP fundamental wave (FW), strong THG at ~420 nm can be observed, corresponding to strong RCP light absorption at ~1260 nm because our RCM demonstrates selective absorption on CPL as shown in Fig. 2e-g; while under LCP pumping, the THG signal can be neglected. The THG-CD can be expressed as:

$$\text{THG-CD} = I_{LCP}^{3\omega} - I_{RCP}^{3\omega} \qquad (1)$$

where $I_{LCP}^{3\omega}$ and $I_{RCP}^{3\omega}$ are emitted intensity of the THG signal of LCP and RCP light. The theoretical limit of CD is 1. Further, the nonlinear g-factor can be expressed as:

$$g_{NL}=2\cdot\frac{I_{LCP}^{3\omega}-I_{RCP}^{3\omega}}{I_{LCP}^{3\omega}+I_{RCP}^{3\omega}} \qquad (2)$$

The THG-CD and nonlinear THG g-factor are also extracted from the simulated and measured data using the above formula, as shown in Fig. 4e and f. A THG-CD value of ~0.92 is experimentally observed at ~420 nm, but in the linear region, the maximum CD is ~0.27, as the inset shown in Fig. 4e; also, the nonlinear g-factor demonstrates a higher value of ~1.8 when compared with its linear one, approaching the theoretical limit 2. This effect can be attributed to the strongly localized near-fields generated by the incident FW (See Supplementary Figs. 14), which produce THG light and coupled with the chiral surface plasmon modes.[26] Since the THG intensity is proportional to the cube of the E-field strength of the illuminating NIR light, this nonlinear relationship can result in a distinct chiroptical response due to the enhanced difference between the THG signals of each polarization. Therefore, the selective harmonic generation for the CP states of the input pump beam, driven by the giant nonlinear circular dichroism, can be utilized to create new functionalities. Plasmonic metasurfaces exhibit strong local field enhancements near resonances, which are much more pronounced for nonlinear interactions. These resonances amplify the nonlinear generation of chiral signals, further increasing the THG-CD compared to linear CD. Moreover, The strength of the nonlinear THG-CD compared to the linear CD can be explained using their underlying mechanisms and mathematical expressions: linear CD originates from the differential absorption of LCP and RCP light. Its intensity is directly proportional to the imaginary part of the chiral susceptibility $\chi^{(1)}$, which describes the material's response to linear optical excitation:

$$\text{Linear CD} = I_{\text{LCP}}-I_{\text{RCP}} \propto \text{Im}(\chi^{(1)}) \qquad (3)$$

The THG process depends on the third-order nonlinear susceptibility $\chi^{(3)}$ in THG-CD. The nonlinear CD is proportional to the difference in THG intensities for LCP and RCP light:

$$\text{THG-CD} = I_{LCP}^{3\omega} - I_{RCP}^{3\omega} \propto \text{Im}(\chi^{(3)}) \qquad (4)$$

The nonlinear susceptibility $\chi^{(3)}$ is generally much larger in magnitude compared to $\chi^{(1)}$ due to resonant enhancement and the involvement of higher-order multipolar interactions. These interactions, amplified by plasmonic effects, significantly boost the response. Since THG is highly sensitive to symmetry and local field enhancements in chiral plasmonic structures, the differential response (numerator) increases more than the total response (denominator), leading to a stronger g-factor than linear optics, as shown in eq. (2).

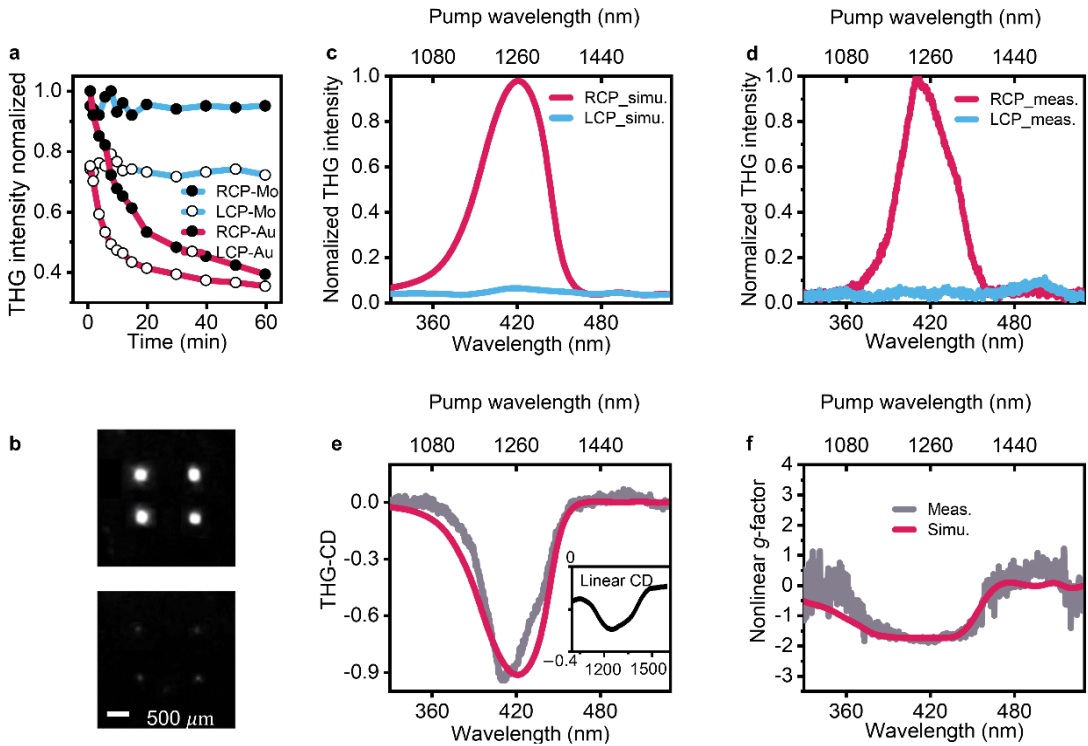

**Fig. 4 Third harmonic generation of the RCM. a**, THG intensity the RCM with $l$=420 nm, and gold control group under 12 GW/cm² excitation. **b**, Microscopy image of THG generated by the RCM ($l$=420 nm) array. **c**. Simulated spectra of normalized THG intensity of the RCM. **d**, Measured spectra of normalized THG intensity of the RCM. **e**, Simulated and measured nonlinear THG-CD; the inset is normalized linear CD spectrum. **f**, Simulated and measured nonlinear THG g-factor.

We further investigate the phenomenon related to the chiral THG of the RCM under CPL light. The input $P_{in}(\omega)$ and the TH signal $P_{out}(3\omega)$ are measured at an FW of $\lambda \approx 1260$ nm. At lower excitation power, the cubic law can be observed while we increase the power further; a deviation in the cubic law can be observed for higher peak intensities due to reverse saturable absorption, as shown in Fig 5a. As the power increases to ~175 mW, the saturated absorption turns into reverse saturable absorption for the RCP excitation. This phenomenon can be explained by the singlet four-level model.[34] This saturated absorption-reverse saturable absorption combined effect may be used in various optical logic gates, such as the XOR gate of the all-optical full-adder.[35] One can couple plasmonic metasurface to intersubband transition to enhance the THG generation efficiency.[24,29] Our chiral metasurface can generate ~500 μW TH signal when pumped at high laser intensity, comparable to those coupled to intersubband transitions, overcoming the drawback of the plasmonic metasurface.[29] The enhanced TH signal can be attributed to intrinsic nonlinear properties of Mo, enhanced electric field, and multi-photon absorption. Also, its conversion efficiency significantly increases at high laser intensity, surpassing its dielectric metasurface with the same pumping intensity level.[36] Based on the THG measurement data in Fig. 5a and $\eta = \frac{P_{out}(3\omega)}{P_{in}(\omega)}$, Fig. 5b shows the THG conversion as an increasing input power. Saturated absorption and reverse saturable absorption can be observed in the conversation efficiency. The shift from saturated absorption to reverse saturable absorption can also be reflected in absorption spectra, as shown in Fig. 5c. One can amplify the chiroptical effect of the RCM by exciting it in the nonlinear region, as the arrows shown in Fig. 5c.

Since our RCM has good thermal stability and demonstrates nonlinear plasmonic chiral effects, we demonstrate a reflective CPL limiter based on the RCM as a proof of concept, as shown in Fig. 5d. The inset demonstrates its working principle. In the nonlinear reverse saturable absorption region, the structure's absorption of RCP increases, and the reflected RCP intensity decreases; the absorbed RCP is converted into heat, and due to its good thermal stability, the structure can be used as a limiter of RCP. The THG effect limits the output of high-intensity light by converting part of the incident light energy into third harmonics, thereby reducing the intensity of the original frequency light. Additionally, nonlinear absorption and scattering induced by high-intensity light further contribute to the operation of the optical limiter. These combined nonlinear effects enable the optical limiter to protect optical systems and sensors under high-intensity light conditions. This discussion is also applicable to LCP illumination. To stress our structure's excellent nonlinear chiroptical effect, we further compare our RCM with other state-of-the-art nonlinear plasmonic structures for THG, as shown in **Table 1**. This is the first metasurface to exhibit a nonlinear chiral g-factor approaching the theoretical limit under high-intensity circularly polarized light pumping. Furthermore, our RCM achieves near-theoretical nonlinear chiroptical responses without additional media (such as quantum well or complex 3D structure), making it an ideal material platform for applications such as nonlinear chiral sensing, optical limiters, and all-optical logic gates.

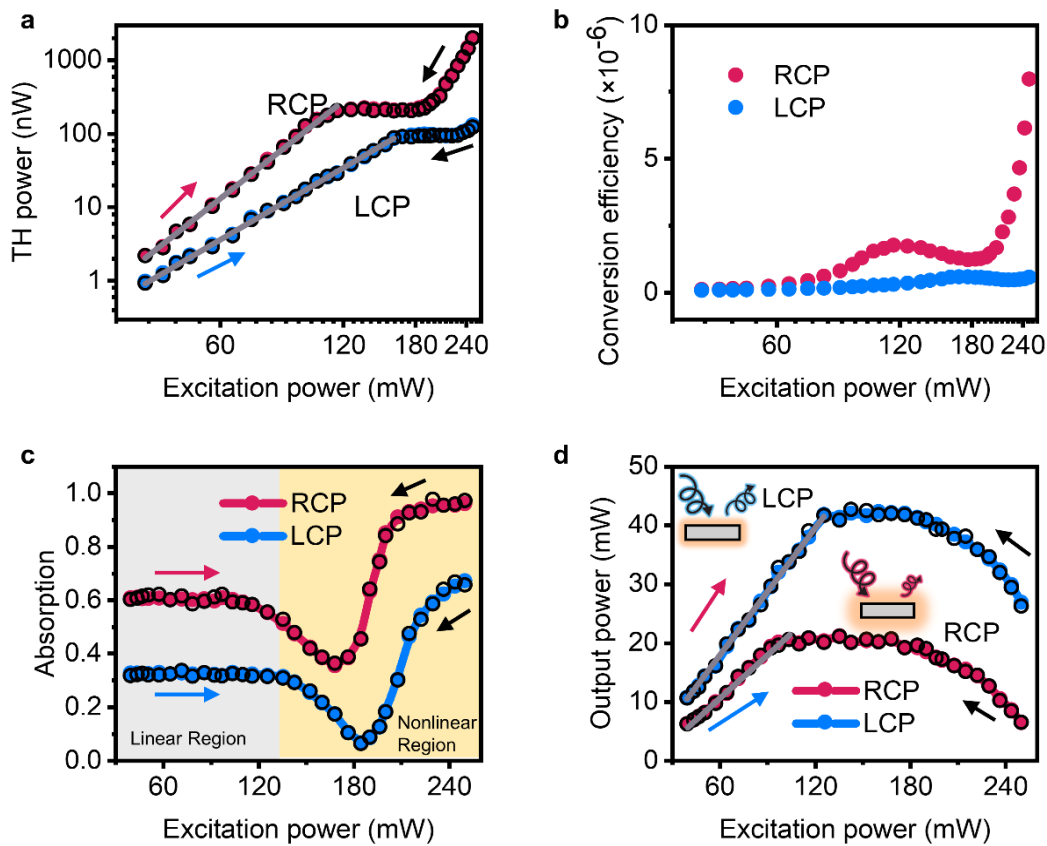

**Fig. 5 Saturated absorption-reverse saturable absorption in RCM. a**, The third harmonic power as a function of the pump power (log-log scale). The filled and hollow circles represent the gradual increase or decrease in pump power; the gray fitted line represents the cubic law. **b**, TH

conversation efficiency as a function of the pump power. **c**, Absorption spectra of the RCM under RCP and LCP light as a function of pump power. **d**, a reflective nonlinear optical limiter based on the RCM. The upper left inset presents the principle of reflective optical limiter. The gray line represents the linear region.

**Table 1. Selected experimental works of nonlinear chiral metastructure.**

| NO. | Nonlinear Medium | Mode | Max. THG-CD or SHG-CD | Max. $g_{NL}$ | FW working wavelength (nm) | Ref. |
|---|---|---|---|---|---|---|
| 1 | AlGaAs+metasurface | T | ~0.65 | / | 1400-1600 | [23] |
| 2 | Au metasurface | T | / | 1.5 | 1000-1100 | [37] |
| 3 | Au metasurface+Multiple quantum well | R | / | 1.96 | 9200-10800 | [24] |
| 5 | Gold helicoid-III nanoparticle | T |  | 1.5 | 600-800 | [26] |
| 6 | Silver metasurface | T | 0.89 | 1.88 | 720-1040 | [38] |
| 7 | Cu and Au 3D strcuture | R |  | 1.64 | 800 | [39] |
| 8 | Only Mo metasurface | R | 0.95 | 1.94 | 1000-1600 nm | This paper |

Note: all values are taken positive for easy comparison. CD and $g_{NL}$ are unified formulas (defined in the main text) for all publications. T and R are reflection mode and reflection mode, respectively.

## Conclusions

In summary, we present a refractory Mo metasurface based on chiral nanostructure. The metasurface not only maintains excellent chiroptical properties but also withstands high temperature and laser power, with survival up to 1100 °C and 12 GW/cm$^2$. Besides, it allows a nonlinear plasmonic effect, demonstrating chiral saturated absorption and reverse saturable absorption. The chiroptical effect, such as CD and g-factor, can be enhanced by exciting the nonlinear effect of the refractory chiral Mo metasurface via increasing pump laser intensity. Our results show a proof-of-concept for optical limiting using the refractory chiral Mo metasurface due to excellent thermal stability and plasmonic chiral effect. Our refractory chiral Mo metasurface can be used not only for reflective optical limiter but also potential for nonlinear chiral sensing, holography, imaging,[38] and laying the foundation towards a material platform for broadband applications such as nonlinear hot electron generation, XOR gate of the all-optical full-adder, and thermoplasmonics. Regarding the plasmonic sensor, the intensity difference of the third harmonic intensity spectrum is more pronounced than its linear counterpart.[20] Conventional metallic materials are damaged under intense lasers, but our structure can provide a research platform for nonlinear hot electron generation. Under nonlinear regions, the generation of hot electrons is probably significantly affected by multi-photon absorption, photoionization, and field enhancement effects.[40]

## Methods

**Device fabrication and characterization**

The Mo thin films are grown using RF magnetron sputtering, with Argon (flow rate: 10SCCM) atmosphere and 50W RF power as the key process parameters. The refractive index of Mo is Measured by a Woollam-M2000D ellipsometer. First, 200 nm Mo film was deposited onto the silicon substrate using a 3-nm Cr sticking layer, followed by 140 nm $SiO_2$ deposition. The top refractory chiral Mo nanostructures are fabricated by an electron beam lithography process using an electron beam lithography system (Elionix ELS-100T) and a 200nm thick spin-coated ZEP520A resist. After the electron beam is exposed to the resist, a development and Mo film deposition process is performed. After, a standard lift-off process is used to yield the structure shown in Fig. 1b. The morphology of the chiral Mo metasurface is characterized by scanning electron microscopy (SEM, SU9000 operated at 30 kV and ZEISS at 5kV).

**Optical simulation**

In this paper, simulations are conducted using COMSOL Multiphysics software through finite element modeling. The electromagnetic properties of the structure are determined by solving Maxwell's equations using the electromagnetic wave frequency domain module. This approach allows for the study of optical absorption, electric and magnetic field distributions, charge distribution, and more. The absorptance (A) is calculated as A=1−R−T, where R represents the reflectance of the structure and T=0. To simulate circularly polarized light CPL, we set up two orthogonally polarized plane waves with a phase difference of 90°. The light is directed perpendicularly onto the upper surface of the chiral metasurface. The surrounding medium is selected to be air. The refractive index of Mo used in the simulation is derived from our experimental data, as illustrated in Supplementary Figs. 8. In the nonlinear simulation, the TH intensity $I_{THG}(3\omega)$ radiated into the far-field is calculated as:

$$I_{THG}(3\omega) \sim |E_{THG}(3\omega)|^2 \sim |\omega \cdot P_{NL}(3\omega)|^2$$

where $E_{THG}(3\omega)$ and $P_{NL}(3\omega)$ are electric field and third order nonlinear polarization term (see Supplementary section for details).

**Optical characterization**

The linear optical absorption and CD spectrum are obtained using a broadband white-light source and a custom infrared microscope coupled with a grating spectrometer. CPL was generated using a polarizer in combination with a quarter waveplate. The quality of the CPL was verified with an additional polarizer. A silver mirror was used as a reference to measure the reflection spectra. The reflection from the front side of the wafer was measured using an unpatterned silicon wafer and then subtracted from the reflection of the patterned Mo metasurface. This setup matches the configuration used in the simulations. The transmission is zero due to the optically thick silver backplane. The schematic of the experimental setup is shown in Supplementary Figs. 15. the light source was replaced with a Ti-sapphire laser source for nonlinear optical absorption measurements. For the nonlinear THG spectroscopy, we employed a 200 fs-pulsed Ti-sapphire laser to pump an optical parametric oscillator (OPO), enabling wavelength-dependent analysis of the RCM. The laser power is adjusted to ~86 mW using an optical power meter (OPM). The sample is adjustable in two directions, enabling polarized illumination and precise laser beam positioning. The CPL beam was directed onto the metasurface using a long-pass (LP) filter and a collimating lens. The harmonic signals emitted from the chiral metasurface were guided to a spectrometer via the long-pass filter and a Bk7 lens. To eliminate the reflected portion of the input pump beam, a short-pass filter (SP) was employed. To analyze the spin state of the output signal, an output quarter-wave plate (QWP) and a linear polarizer were placed in front of the spectrometer. The optical setup is schematically shown in Supplementary Figs. 16.

## Acknowledgments

Peng Yu and Tianji Liu contribute equally to this work. Peng Yu acknowledges the funding supported by the Sichuan Science and Technology Program 2024NSFSC0483 and 2024JDHJ0033. Tianji Liu acknowledges the funding supported by the Natural Science Foundation of Jilin Province (No.20240101314JC). A part of this work was supported by the "Nanotechnology Platform Project (Nanotechnology Open Facilities in Osaka University)" of the Ministry of Education, Culture, Sports, Science and Technology, Japan (No. F-19-OS-0004).

## Competing interests

The authors declare no other competing interests.